\documentclass[12pt,oupdraft]{bio}
\usepackage{url}
\usepackage{comment}
\usepackage{tabularx}
\usepackage{graphicx}
\usepackage{adjustbox}
\usepackage{chngcntr}
\counterwithout{equation}{section}
\usepackage{footnote}
\makesavenoteenv{tabular}
\makesavenoteenv{table}
\usepackage{booktabs,caption}
\usepackage[flushleft]{threeparttable} 
\usepackage{emptypage}
\usepackage{chadstyle}
\usepackage{fancyhdr}
\usepackage{geometry}
\usepackage{mathtools}

\begin{document}

\title{Sensitivity Analysis of Causal Treatment Effect Estimation for Clustered Observational Data with Unmeasured Confounding}

\author{Yang Ou$^1$, Lu Tang$^1$, Chung-Chou H. Chang$^{1,2}$\\[4pt]
\textit{$^{1}$Department of Biostatistics, Graduate School of Public Health, University of Pittsburgh, Pittsburgh, Pennsylvania \break
$^{2}$Department of Medicine, School of Medicine, University of Pittsburgh, Pittsburgh, Pennsylvania}
 }

\maketitle 
\begin{abstract}
{Identifying causal treatment (or exposure) effects in observational studies requires the data to satisfy the unconfoundedness assumption which is not testable using the observed data. With sensitivity analysis, one can determine how the conclusions might change if assumptions are violated to a certain degree. In this paper, we propose a new technique for sensitivity analysis applicable to clusters observational data with a normally distributed or binary outcome. The proposed methods aim to assess the robustness of estimated treatment effects in a single study as well as in multiple studies, i.e., meta-analysis, against unmeasured confounders. Simulations with various underlying scenarios were conducted to assess the performance of our methods. Unlike other existing sensitivity analysis methods, our methods have no restrictive assumptions on the number of unmeasured confounders or on the relationship between measured and unmeasured confounders, and do not exclude possible interactions between measured confounders and the treatment. Our methods are easy to implement using standard statistical software packages.}
{Sensitivity analysis; Unmeasured confounding; Bias factor; Clustered observational data; Conditional average treatment effect; Mixed model}
\end{abstract}

\newpage
\section{Introduction}
When attempting to estimate the causal effect of an experimental treatment, a risk exposure, or a new policy, researchers often prefer the approach of randomized experimental design. Through randomization, patients who are assigned to different treatment arms are comparable in baseline factors thus potential biases can be avoided. However, when it is not feasible to conduct randomized experiments because of ethical, financial, or other concerns, observational studies are brought into play to estimate the casual treatment effect. To identify the average treatment effects (ATE) and the conditional average treatment effects (CATE) from observational data, further assumptions must be made. The no unmeasured confounder assumption, which is also referred to as the exchangeability assumption, is one of the assumptions needed to identify ATE and CATE from observational data which asserts that the confounding mechanism is determined by observed covariates only. One must address the concern of bias caused by the violation of the exchangeability assumption. The bias introduced by unmeasured treatment-outcome confounding will threaten the usefulness of the statistical inferences that rely on this assumption. For example, \cite{Willett1990} claimed dietary carotenoids reduce the risk of lung cancer using observational data, while a randomized clinical trial conducted later (\citealp{group}) showed that the group that received carotene has a higher incidence of lung cancer than those that did not receive carotene. The contradictory conclusions are most likely because that the observational study was unable to adjust for unmeasured confounders. For example, people taking dietary carotenoids tend to be healthier due to social and behavioral factors that were beneficial to them. However, adjusting for measured confounders only was not enough to adjust for the difference of health condition between groups taking and not taking carotenoids. Similar contradictory conclusions have been seen in observational and randomized studies analyzing the relationship between antioxidant vitamins and cardiovascular disease (\citealp{Stampfer1993}, \citealp{Gaetano2001}, \citealp{Lawlor2004}), which could also be results of violation of the exchangeability assumption.

To deal with unmeasured confounding, researchers propose to use instrumental variable analysis with an exogenous variable, i.e., an \textit{instrument} that is not directly related to the outcome yet can affect the choice of treatments. This method uses an approximated randomization to form comparison groups of similar characteristics including both measured and unmeasured confounders. 
	
The difference-in-differences (DID) approach can be used for estimating ATE in the presence of unobserved confounding when pre- and post-treatment outcome measurements are available. This approach requires that the association between the unobserved confounder and the outcome is equal in the two treatment arms and that it is a constant over time (\citealp{Sofer}).  

Another way to account for unmeasured confounder is through the use of a proxy variable. \cite{Tchetgen} introduced \textit{negative controls} which are auxiliary variables not causally associated with the treatment or outcome, and a generalization of negative controls, namely the \textit{proximal causal learning framework}, that offers an opportunity to learn causal effects in settings which the assumption of no unmeasured confounding fails. 

Among all methods dealing with unmeasured confounders in observational studies, the most common approach is using sensitivity analysis to assess the robustness of causal effect estimation and inference against unmeasured confounders, i.e., how much the estimation will be affected by unmeasured confounders. Sensitivity analysis methods for various models and data structures have been proposed in \cite{Lin}, \cite{Rosenbaum}, \cite{Imbens2003}, \cite{Marcus1997}, and \cite{Robins2000}. These methods are applicable under additional restrictions, for examples, only for a single binary type of unmeasured confounder (\citealp{Rosenbaum}); or assuming conditional independence that requires the measured and unmeasured confounders to be independent conditional on the treatment (\citealp{Lin}). \cite{VanderWeele1} and \cite{VanderWeele2} proposed a  sensitivity method to estimate the \textit{E-value} for binary outcome, which is the minimum risk ratio a unmeasured confounder must have with both treatment and outcome to explain away the confounded treatment effect. \cite{Mathur2019} extended the idea by developing a sensitivity analysis method that is applicable to multiple studies (meta-analysis). To estimate the ATE on a continuous outcome, \cite{VanderWeele1} approximated the difference in the mean continuous outcomes between two treatment arms to a risk ratio then applied the E-value on this approximated risk ratio under additional assumptions.

In this study, we propose a general sensitivity analysis technique for normally distributed and binary outcomes without making distributional assumptions for the measured and unmeasured confounders. We successfully incorporated data from clustered observational designs (e.g., hierarchical or longitudinal) without additional restrictions on the number of unmeasured confounders or the interaction effects between treatment and measured confounders. We also extended the idea and developed a method for multiple studies (meta-analysis). 

The rest of the paper is organized as follows. Section \ref{MM_methods} introduces the concept of our proposed methods, including cases for single and multiple studies. Section \ref{MM_interpretation} shows the interpretation of our methods. Section \ref{MM_simulation} provides the assessment of our proposed methods under several simulated scenarios. We conclude our findings and provide final comments in Section \ref{MM_discussion}.

\section{Methods}
\label{MM_methods}
 
\subsection{Single Study - Continuous Outcome} 
Consider a binary treatment variable $A$ with value 1 for the treated and $0$ for the control, and a continuous outcome variable $Y$. Denote $\textbf{X}$ as a set of measured confounders, and $\textbf{U}$ as a set of unmeasured confounders. Without loss of generality, we assume one measured confounder and one unmeasured confounder. Let $Y^{A=a}$ (or $Y^a$ ) be the counterfactual or potential outcome, which is the outcome that would have been observed if individual had been given treatment $a$. Denote $Y$ as the observed outcome that corresponds to the treatment value actually experienced.

Individual causal effects are defined as the difference between potential outcomes. Because  only one of those outcomes is observed for each individual, individual effects cannot be identified. The average treatment effects (ATE) and conditional average treatment effects (CATE) were proposed and can be estimated from the observed data when additional assumptions hold. The causal ATE of $A$ on $Y$ is the difference between the expectation of the potential outcomes in the population while the conditional causal ATE is the ATE in any subpopulation. \begin{gather*} 
\text{ATE}=\mathbb{E}\left(Y^{A=1}\right)-\mathbb{E}\left(Y^{A=0}\right), \\
\text{CATE}=\mathbb{E}\left(Y^{A=1} \mid X=x, U=u\right)-\mathbb{E}\left(Y^{A=0} \mid X=x, U=u\right) .
\end{gather*}
To estimate the ATE or CATE, the following causal assumptions are necessary. The stable unit treatment value assumption (SUTVA) indicates that treatment assignment of one subject does not affect the outcome of another subject. The consistency assumption requires that the potential outcome under treatment $A=a$, $Y^{A=a}$, is equal to the observed outcome if treatment $A=a$ is actually received, i.e.,
$
Y^{a}=Y \text{ if } A=a, \text{ for all } a.
$
The positivity assumption states that treatment assignment is not deterministic for every set of values of $X$ and $U$, i.e.,  
$
0<P(A=a \mid X=x,U=u)<1 \text{ for all } a \text{, } x\text{ and }  u.
$
The exchangeability assumption requires that given $X$ and $U$, treatment assignment is independent of the potential outcomes, i.e.,
$Y^{a} \perp A \mid X,U$ for all $a$. 

Under these assumptions, ATE and CATE can be estimated from the data collected from an observational study:
\begin{equation*}
\begin{split}
&\mathbb{E}\left(Y^{A=1} \mid X=x,U=u\right) - \mathbb{E}\left(Y^{A=0} \mid X=x,U=u\right) \\&= \mathbb{E}\left(Y^{A=1} \mid A=1,X=x,U=u\right) - \mathbb{E}\left(Y^{A=0} \mid  A=0,X=x,U=u\right)\\
&=\mathbb{E}\left(Y \mid A=1,X=x,U=u\right) - \mathbb{E}\left(Y \mid  A=0,X=x,U=u\right).
\end{split}
\end{equation*} 
 
In clustered observational study, a common approach to estimate $\mathbb{E}\left(Y \mid A, X=x,U=u\right)$ is to use a generalized mixed effects model or a generalized linear model with GEE, where all potential confounders are included. We will introduce our sensitivity analysis method based on generalized mixed effects model, where random effects will be used to account for within-cluster correlation.  Let $i$ $(i=1, \dots, I)$ and $j$ $(j=1, \dots, J)$ be the indicators for the level-1 (e.g., subjects nested in clusters) and the level-2 unit (e.g., clusters), respectively. Consider a continuous outcome $Y_{ij}$ with the marginal mean model:
\begin{equation*}
\small \begin{split} 
\mathbb{E}(Y_{ij}|A_{ij},X_{ij},U_{ij})=\beta_{0} + \beta_{1} A_{ij}+\beta_{2} X_{ij}+ \beta_{3} A_{ij} X_{ij}+\theta U_{ij};
\end{split}
\end{equation*}  
and the corresponding mixed model: 
\begin{equation}
\small \begin{split} \label{1}
&Y_{ij}=\beta_{0} + \beta_{1} A_{ij}+\beta_{2} X_{ij}+ \beta_{3} A_{ij} X_{ij}+\theta U_{ij}+\zeta_j+\epsilon_{ij},
\end{split}
\end{equation}  
where $\boldsymbol \beta=(\beta_0,\beta_1,\beta_2,\beta_3)^T$ and $\theta$ are unknown regression parameters; $\zeta_{j}$ and $\epsilon_{ij}$ are the random intercepts and the level-1 error terms, respectively, where $\zeta_{j} \sim N(0,\nu)$ and $\epsilon_{ij}| \zeta_{j}\sim N(0,\phi)$ with unknown variances $\nu$ and $\phi$.

Our goal is to estimate the CATE, denoted as $E_x$, with the form
\begin{equation*}
\begin{split}
E_x  \coloneqq  \mathbb{E}\left(Y \mid A=1,X=x,U=u\right) - \mathbb{E}\left(Y \mid  A=0,X=x,U=u\right) =\beta_1+ x\beta_3.
\end{split}
\end{equation*} 
Since $U$ is unmeasured, we can only fit the following confounded model:
\begin{equation}  \label{2}
Y_{ij}=\beta_{0}^{c} + \beta_1^c A_{ij}+\beta_2^c X_{ij}+ \beta_3^c A_{ij} X_{ij}+\zeta_j^c+\epsilon_{ij}^c,
\end{equation}
where $\boldsymbol \beta^{c}=(\beta_0^{c},\beta_1^{c},\beta_2^{c},\beta_3^{c})^T$ are unknown regression parameters; $\zeta_{j}^c$ and $\epsilon_{ij}^c$ are random intercepts and level-1 error terms, respectively, where $\zeta_{j}^c \sim N(0,\nu^c)$ and $\epsilon_{ij}^c| \zeta_{j}^c\sim N(0,\phi^c)$ with unknown variances $\nu^c$ and $\phi^c$. There is potential violation of assuming a constant variance of random effect and error term, but our methods are robust against this potential violation based on the simulation results. 

Denote confounded treatment effect $E_x^c$ as:
\begin{equation*} 
E_x^c \coloneqq  \mathbb{E}\left(Y \mid A=1,X=x \right) - \mathbb{E}\left(Y \mid  A=0,X=x \right) =\beta_1^c+ x\beta_3^c. 
\end{equation*} 
We refer $\boldsymbol \beta $ and $\boldsymbol \beta^{c} $ as the true and confounded coefficients, respectively. The element in $\boldsymbol \beta $ cannot be estimated directly using the observed data while $\boldsymbol \beta^{c} $ can. Therefore, our goal is to establish the relationship between $\boldsymbol \beta $ and $\boldsymbol \beta^{c} $ to infer $\boldsymbol \beta $.   

Let $F(U | A, X)$ be the cumulative distribution function of $U$ given $A$ and $X$. The conditional expectation of $Y_{ij}$ given $A_{ij}$ and $X_{ij}$ based on model (\ref{1}) becomes: 
\begin{equation} \label{3}
\begin{split} 
\mathbb{E}(Y_{ij}| A_{ij}, X_{ij}=x)&= \mathbb{E_{U}}{ \mathbb{E}_{\zeta} \mathbb{E}(Y_{ij}| A_{ij}, X_{ij}=x,U_{ij},\zeta_j)  }\\
&= \mathbb{E_{U}}  \mathbb{E}(Y_{ij}| A_{ij}, X_{ij}=x,U_{ij},\zeta_j=0)  \\
&=\beta_0 + \beta_1 A_{ij} +x\beta_2 +x\beta_3 A_{ij}+\theta \int_{-\infty}^{\infty}U_{ij} d F(U_{ij} | A_{ij}, X_{ij}=x)\\
&=(\beta_0+ \theta m_{0x})  +  \{ \beta_1+ x\beta_3+ \theta  (m_{1x}- m_{0x} )\}A_{ij}+x\beta_2, 
\end{split}
\end{equation}
where $m_{1x}=\mathbb{E}(U_{ij}| A_{ij}=1, X_{ij}=x)$ and $m_{0x}=\mathbb{E}(U_{ij}| A_{ij}=0, X_{ij}=x)$.

\noindent The conditional expectation of $Y_{ij}$ based on the confounded model described in equation (\ref{2}) has the form:
\begin{equation} \label{4}
\mathbb{E}(Y_{ij}| A_{ij}, X_{ij}=x)=\beta_{0}^{c} + (\beta_{1}^{c} + x\beta_{3}^{c})A_{ij} +x\beta_{2}^{c}. 
\end{equation}

\noindent Comparing equations (\ref{3}) and (\ref{4}), we have the following relationship:
\begin{equation} \label{bias1}
E_x=E_x^c -B_x, 
\end{equation}
where $E_x=\beta_1 + x\beta_3$, $E_x^c=\beta_1^c + x\beta_3^c$, and $B_x=(m_{1x}-  m_{0x})\theta$. Note that $E_x$ is the CATE conditional on $X =x$ and $E_x^c$ is the confounded treatment effect conditional on $X =x$. Equation (\ref{bias1}) specifies the relationship between causal CATE and confounded treatment effect. Define $B_x$ as the bias factor, which is a function of the effect of unmeasured confounders and the difference of the mean unmeasured confounders between the treated and control groups given $X=x$. Let $\theta$, $m_{1x} $ and $m_{0x}$ be the sensitivity parameters.

\subsection{Sensitivity Analysis for Single Study with Continuous Outcome}
\label{interpretation} 
Denote the estimated confounded treatment effect as $ \widehat{E}_x^c$ (95\% CI, $lb$ to $ub$), where $lb$ and $ub$ are the lower and upper bounds of its 95\% confidence interval. Following equation (\ref{bias1}), the estimated conditional average treatment effects is $ \widehat{E}_x^c-B_x$ (95\% CI, $lb-B_x$ to $ub-B_x$). We say an unmeasured confounder can explain away the estimated confounded treatment effect if the unmeasured confounder is able to shift the confidence interval of the estimated confounded treatment effect to include the null value (no treatment effect). 
If confidence interval of estimated confounded treatment effect includes the null value, then no confounding is needed to explain away the results or move confidence interval to include the null value.

Once the sensitivity parameters are specified, we can calculate bias factor and the estimated CATE. However, in practice, we do not know the strengths and distributions of unmeasured confounders, thus, the sensitivity parameters are usually unknown. What we can do is to calculate how the estimate is affected under different values of the bias factor. Among all different bias factors that are able to explain away the confounded treatment effect, \textit{minimal bias factor} is defined as the smallest value of bias factors that is able to explain away the confounded treatment effect, meaning any bias factor larger than the minimal bias factor can also explain away the confounded treatment effect. The robustness of the estimated confounded treatment effect can be measured by the minimal bias factor. A small value of the minimal bias factor indicates the estimated treatment effect is not robust to unmeasured confounders. 

Let $\widehat{E}_x^c>0$ and $\widehat{E}_x^c<0$ be apparently positive effect and apparently negative effect, respectively. In apparently positive effect case, the confounded treatment effect will be explained away when the lower bound of confidence interval of estimated confounded treatment effect is smaller than 0, i.e., $lb- {B }_x \le  0$. It means any bias factor, $ B_x $, such that $ B_x \ge lb$, can explain away the confounded treatment effect. Denote $B_x^*$ as the minimal bias factor that can explain away the confounded treatment effect. Thus, when $\widehat{E}_x^c>0$, the minimal bias factor that can explain away the confounded treatment effect is $ lb$, i.e., $ {B}_x^*= lb$. 

In the apparently negative case, define $\Tilde{B }_x=   -{B }_x $, thus, estimated CATE can be expressed as $ \widehat{E}_x^c+\Tilde{B }_x$ (95\% CI, $lb+\Tilde{B }_x$ to $ub+\Tilde{B }_x$). 
The confounded treatment effect will be explained away when $ub+\Tilde{B }_x \ge  0$. It means any bias factor, $ \Tilde{B }_x$, such that $ \Tilde{B }_x \ge -ub$, can explain away the confounded treatment effect. Denote $\Tilde{B}_x^*$ as the minimal bias factor that can explain away the confounded treatment effect. 
Thus, when $\widehat{E}_x^c<0$, the minimal bias factor that can explain away the confounded treatment effect is $ -ub$, i.e., $ \Tilde{B}_x^*= -ub$.
 
As shown above, bias factor is a function of the effects of unmeasured confounders, and the difference of mean of unmeasured confounders between treated and control groups given $X=x$, such that $B_x=(m_{1x}-  m_{0x})\theta$ and $\Tilde{B }_x=-(m_{1x}-  m_{0x})\theta$. It means that, when $\widehat{E}_x^c>0$, a set of $\theta$, $m_{1x} $, and $m_{0x}$, such that $ (m_{1x}-  m_{0x})\theta \ge lb$, can explain away estimated confounded treatment effect. When $\widehat{E}_x^c<0$, a set of $\theta$, $m_{1x} $, and $m_{0x}$, such that $ (m_{0x}-  m_{1x})\theta \ge -ub$, can explain away estimated confounded treatment effect. We recommend to provide a table or contour plot which present different values of $\theta$, $m_{1x} $, $m_{0x}$, and bias factor to give researchers a sense of what are combinations of sensitivity parameters that can explain away the confounded treatment effect and therefore a sense of how sensitive the confounded treatment effect is to unmeasured confounders.
  
\subsection{Single Study - Binary Outcome} 
For a binary outcome, risk difference (RD) and relative risk (RR) can be used to explore causal treatment effect. In this article, RR will be used to explore the causal relationship. With same notations as in the case of continuous outcome, except that $Y$ denotes a binary variable now, when same assumptions hold: 
\begin{equation*}
\begin{split}
\text{RR}=\frac{\mathbb{E}\left(Y^{A=1} \mid X=x,U=u\right)}{\mathbb{E}\left(Y^{A=0} \mid X=x,U=u\right)}   = \frac{\mathbb{E}\left(Y^{A=1} \mid A=1,X=x,U=u\right)}{\mathbb{E}\left(Y^{A=0} \mid  A=0,X=x,U=u\right)} =\frac{\mathbb{E}\left(Y \mid A=1,X=x,U=u\right)}{\mathbb{E}\left(Y \mid  A=0,X=x,U=u\right)} .
\end{split}
\end{equation*} 
When the event is rare in both treatment and control group, it can be shown that odds ratio is approximately equal to the relative risk. A logistic regression model is widely used to estimate odds ratio. Let the response variable $Y$ be related to $A$, $X$, and $U$ through following generalized linear mixed model:
\begin{equation} \label{binaryall}
\text{logit}\left\{\mathbb{P}(Y_{ij}=1|A_{ij},X_{ij},U_{ij},\zeta_{j})\right\}=\beta_{0} + \beta_{1} A_{ij}+\beta_{2} X_{ij}+ \beta_{3} A_{ij} X_{ij}+\theta U_{ij}+\zeta_{j},
\end{equation}  
where $\boldsymbol \beta=(\beta_0,\beta_1,\beta_2,\beta_3)^T$ and $\theta$ are unknown regression parameters, and $\zeta_{j} \sim N(0,\nu )$. 

When $v  $ is small, it can be shown (proof is provided in Appendix \ref{binaryoutcomeA}) 
\begin{equation*}   
 \text{logit}\{\mathbb{P}(Y_{ij}=1| A_{ij}, X_{ij},U_{ij} )\}\approx \beta_{0} + \beta_{1} A_{ij} +\beta_{2} X_{ij} + \beta_{3} A_{ij}  X_{ij}+\theta U_{ij}.
\end{equation*} 
Denote CATE on the log-RR scale as $E_x $. Thus, when the event is rare and $v$ is small, $E_x $ has the form:
$ 
E_x \coloneqq \text{log}\frac{\mathbb{E}\left(Y^{A=1} \mid X=x,U=u\right)}{\mathbb{E}\left(Y^{A=0} \mid X=x,U=u\right)}   
 =\text{log}\frac{\mathbb{E}\left(Y \mid A=1,X=x,U=u\right)}{\mathbb{E}\left(Y \mid  A=0,X=x,U=u\right)} 
 \approx \beta_1+ x\beta_3. 
$

Since $U$ is unmeasured, we can only fit the following confounded model:
\begin{equation} \label{binaryreduce}
\text{logit}\left\{\mathbb{P}(Y_{ij}=1|A_{ij},X_{ij},\zeta_{j}^c)\right\}=\beta_{0}^c + \beta_{1}^c A_{ij}+\beta_{2}^c X_{ij}+ \beta_{3}^c A_{ij} X_{ij} +\zeta_{j}^c,
\end{equation}
where $\zeta_{j}^c \sim N(0,\nu^c)$ and $\boldsymbol \beta^c=(\beta_0^c,\beta_1^c,\beta_2^c,\beta_3^c)^T$ are unknown regression parameters. 
Similarly, when $\nu^c$ is small, it can be shown that for the confounded model:
\begin{equation} \label{binaryf2}
\text{logit}\left\{\mathbb{P}(Y_{ij}=1| A_{ij}, X_{ij} )\right\}\approx \beta_{0}^c + \beta_{1}^c A_{ij} +\beta_{2}^c X_{ij} + \beta_{3}^c A_{ij}  X_{ij}. 
\end{equation}
Denote confounded treatment effect on the log-RR scale as $E_x^c$ which has the form:
$
E_x^c\coloneqq \text{log}\frac{\mathbb{E}\left(Y \mid A=1,X=x\right)}{\mathbb{E}\left(Y \mid  A=0,X=x\right)} \approx \beta_1^c+ x\beta_3^c. 
$
 
When $U$ is binary and both $\theta$ and $\nu$ are small, we have (proof is provided in Appendix \ref{binaryoutcome}): %
\begin{equation} \label{binaryf3} \small
\begin{split}
&\text{logit}\left\{\mathbb{P}(Y=1| A, X=x)\right\}\approx \beta_{0} + \beta_{1} A +\beta_{2} x + \beta_{3} A  x+  \text{log}\{P_{Ax}  \text{exp}(\theta +\frac{\nu}{2 })+ (1-P_{Ax})  \text{exp}( \frac{\nu}{2 }) \}\\
&=\beta_{0} +\beta_{2} x+\text{log}\{P_{0x}  \text{exp}(\theta +\frac{\nu}{2 })+ (1-P_{0x})  \text{exp}( \frac{\nu}{2 }) \}+ \{\beta_{1} + \beta_{3} x+\text{log}\frac{P_{1x}  \text{exp}(\theta )+ 1-P_{1x}}{P_{0x}  \text{exp}(\theta )+ 1-P_{0x}} \} A,
\end{split}
\end{equation}  
where $P_{AX}=P(U_{ij}=1 | A_{ij}, X_{ij})$.

Following the confounded model described in equation (\ref{binaryf2}), we have:
\begin{equation} \label{binaryf4}
\text{logit}\left\{\mathbb{P}(Y =1| A , X =x )\right\}\approx \beta_{0}^c + \beta_{1}^c A_{ij} +\beta_{2}^c x + \beta_{3}^c A   x =\beta_{0}^c +\beta_{2}^c x+ (\beta_{1}^c +\beta_{3}^c x)A .
\end{equation}

When $\theta$, $\nu$, and $\nu^c$ are small, comparing equations (\ref{binaryf3}) and (\ref{binaryf4}), we have the following relationship:
\begin{equation} \label{bias2}
E_x=E_x^c -B_x, 
\end{equation}
where $E_x=\beta_1 + x\beta_3$, $E_x^c=\beta_1^c + x\beta_3^c$, and $B_x=\text{log} \frac{P_{1x}  \text{exp}(\theta )+ 1-P_{1x}}{P_{0x}  \text{exp}(\theta )+ 1-P_{0x}}$.

When $U$ is normally distributed such that $U\sim N(\mu_{AX},\sigma_u)$, i.e., the expectation of $U$ is dependent on both $A$ and $X$, then if $\theta$, $\sigma_u$, and $\nu $ are small, we have (proof is provided in Appendix \ref{binaryoutcome2}): %
\begin{equation} \label{normal3}
\begin{split}
\text{logit}\left\{\mathbb{P}(Y=1| A, X=x)\right\}&\approx \beta_{0} + \beta_{1} A +\beta_{2} x + \beta_{3} A  x+  \theta\mu_{Ax} +\frac{\theta^2\sigma_u+\nu}{2 }\\
&=\beta_{0} +\beta_{2} x+\mu_{0x}  +\frac{\theta^2\sigma_u+\nu}{2 }+ (\beta_{1}+    x\beta_{3}+\theta\mu_{1x}-\theta\mu_{0x} )A.  
\end{split}
\end{equation}    
When $\theta$, $\sigma_u$, $\nu$, and $\nu^c$ are small, comparing equations (\ref{binaryf4}) and (\ref{normal3}), we have the following relationship:
\begin{equation} \label{bias3}
E_x=E_x^c -B_x, 
\end{equation}
where $E_x=\beta_1 + x\beta_3$, $E_x^c=\beta_1^c + x\beta_3^c$, and $B_x=\theta(\mu_{1x}-  \mu_{0x})$. 

Intraclass correlation coefficient (ICC) is the proportion of total variance of the outcome that is attributable to the between-subject variation, i.e., how much variation of the outcome can be explained by the random effects. For a logistic model, $\text{ICC}=\frac{\nu}{\nu+\pi/3}.$
It means when $\theta$, ICC, and confounded ICC are small, equations (\ref{bias2}) and (\ref{bias3}) hold.

With equations (\ref{bias2}) and (\ref{bias3}), we can calculate the minimal bias factor that can explain away the confounded treatment effect by moving the confidence interval to include the null value. Similarly, if a small minimal bias factor is able to explain away the confounded treatment effect, the confounded treatment effect is not robust to unmeasured confounders.

\subsection{Multiple Studies} 
When multiple observational studies is involved where only study-level estimates not individual-level data can be merged, meta-analysis is the most commonly used analysis technique to estimate the population covariate effects which combines estimates from all studies and takes study heterogeneous into consideration. In this section, we introduce our proposed method to assess how sensitive the estimated population covariate effects from meta-analysis are to the unmeasured confounders.
 
Suppose we have $K$ studies which are random samples from a population of interest. For study $k$, let $Y_{ijk}$ be the continuous response variable, which is related to treatment assignment $A_{ijk}$, measured confounders $X_{ijk}$, and unmeasured confounder $U_{ijk}$ through the following random intercepts model:
$$   \label{ddd}
Y_{ijk}=\beta_{0k} + \beta_{1k} A_{ijk}+\beta_{2k} X_{ijk}+ \beta_{3k} A_{ijk} X_{ijk}+\theta_k U_{ijk}+\zeta_{jk}+\epsilon_{ijk},
$$  
where $i$ $(i=1, \dots, I)$ and $j$ $(j=1, \dots, J)$ are the indicators for the level-1 (e.g.,subjects in clusters) and the level-2 unit (e.g., clusters), respectively; and $\beta_k$'s and $\theta_k$ are the regression parameters adjusting for both measured and unmeasured confounders. Similarly, when causal assumptions hold, the model can be used to estimate the causal ATE. Denote $E_{k|x} $ as the CATE for study $k$ given $X =x$, such that $E_{k|x} =\beta_{1k} +x\beta_{3k}$.

Since $U_{ijk}$ is unmeasured, we can only fit the following confounded model for each of the $K$ studies:
$$ 
Y_{ijk}=\beta_{0k}^{c} + \beta_{1k}^{c} A_{ijk}+\beta_{2k}^{c} X_{ijk}+ \beta_{3k}^{c} A_{ijk} X_{ijk}+\zeta_{jk}^{c}+\epsilon_{ijk}^{c},
$$
where $\beta_k^c$'s are confounded regression parameters only adjusting for measured confounders. Denote $E_{k|x}^c$ as the confounded treatment effect for study $k$ given $X =x$, such that $E_{k|x}^c=\beta_{1k}^{c}+x\beta_{3k}^{c}$.
 
In a standard setup of parametric random effects meta-analysis (\citealp{Borenstein:2007}), it assumes that each of the $K$ studies measures a potentially unique confounded effect size $E_{k|x}^c$, such that $E_{k|x}^c \sim N(\mu_{E_{x}^c},V_{E_{x}^c})$, for $k=1, \dots, K$. Let $ \widehat E_{k|x}^c$ be the point estimate of the treatment effect and $\nu_k $ be the known within-study variance for study $k$. In random effects meta-analysis setup, $ \widehat E_{k|x}^c$ is determined by the true confounded effect size $E_{k|x}^c$ plus the known within-study variance $\nu_k $, while $E_{k|x}^c$ is determined by the grand mean, $\mu_{E_{x}^c}$ and the between-study variance, $V_{E_{x}^c}$. More generally, for any observed effect $\widehat E_{k|x}^c$,
$$
 \widehat E_{k|x}^c= \mu_{E_{x}^c}+\varphi_{k}+\kappa_{k},
$$
where $\varphi_{k} \sim N(0,V_{E_{x}^c})$ and $\kappa_{k} \sim N(0,\nu_k)$. Fixed effects meta-analysis can be viewed as a special case where $V_{E_{x}^c}=0$.


The relationship between the CATE and confounded treatment effects in equation (\ref{bias1}) still hold for each of the $K$ studies as we illustrated in the single study case,
\begin{gather*} \label{bound}
E_{k|x}=E_{k|x}^{c}-B_{k|x}; \hspace{0.5 in} 
B_{k|x}=(m_{k|1x}-  m_{k|0x})\theta_k,
\end{gather*} 
where $m_{k|1x}=\mathbb{E}(U_{ijk}| A_{ijk}=1, X_{ijk}=x)$ and $m_{k|0x}=\mathbb{E}(U_{ijk}| A_{ijk}=0, X_{ijk}=x)$, $B_{k|x} \sim N(\mu_{B_x},V_{B_x})$ and $E_{k|x}^{c} \sim N(\mu_{E_{x}^c},V_{E_{x}^c})$. It means that we allow the bias factor $B_{k|x}$ to vary across studies. Assume that $B_{k|x}$ is independent of $E_{k|x}$. $B_{k|x}$ is independent of $E_{k|x}$ but not $E_{k|x}^{c}$ because studies with a larger bias factor may observe a larger confounded treatment effect. Thus, the distribution of the CATE can be written as $E_{k|x} \sim N(\mu_{E_{x}},V_{E_{x}})$, where
\begin{equation*} \label{meta}
     \mu_{E_{x}} =\mu_{E_{x}^c}-\mu_{B_x}; \hspace{0.5 in} V_{E_{x}}=V_{E_{x}^c}-V_{B_x}. 
\end{equation*}

\subsection{Sensitivity Analysis for Multiple Studies} \label{metasection}  Although we use continuous outcome as an example to demonstrate the method in multiple studies, similar derivation can be easily extended to the binary outcome case by replacing the error distribution and the link function in the generalized linear mixed model for binary outcomes. Given equations (\ref{bias2}) and (\ref{bias3}) hold for each of the $K$ studies, we can similarly derive the distribution of CATE under the case of binary outcomes.

Following the newly designed metrics for meta‐analyses of heterogeneous effects in \cite{Mathur20192} and \cite{Mathur2019}, we define that there is a treatment effect if the probability of observing a CATE larger than a scientifically meaningful effect size $q$, is larger than a threshold $r$. This means that a bias factor will explain away the confounded treatment effect in meta-analysis if the bias factor can reduce the probability of observing a conditional average treatment effects larger than $q$ to less than $r$. Note, for continuous outcome, $q$ is on the difference of the continuous outcomes scale, while for binary outcome, $q$ is on the log-RR scale.

Denote $\widehat \mu_{E_{x}^c} $ and $\widehat V_{E_{x}^c}$ as the estimators of $\mu_{E_{x}^c}$ and $V_{E_{x}^c}$. Let $\widehat\mu_{E_x^{c}}>0$ and $\widehat\mu_{E_x^{c}}<0$ be the apparently positive effect and apparently negative effect, respectively, in meta-analysis.   

For the apparently positive effect, we define the probability of observing a CATE larger than $q$ as $p(q)_x=P\left(E_{k|x} >q\right)$ where $E_{k|x} \sim N(\mu_{E_x},V_{E_x})$. $p(q)_x$ can be expressed using $\mu_{B_x }$, $V_{B_x }$, $ \mu_{E_x^{c}}$, and $ V_{E_x^{c}}$:
\begin{equation} \label{pq}
 {p}(q)_x=1-\Phi\left(\frac{q+\mu_{B_x }- \mu_{E_x^{c}}}{\sqrt{   V_{E_x^{c}} -V_{B_x } }}\right), \quad     V_{E_x^{c}} >V_{B_x } 
\end{equation} 
where $\Phi$ is the cumulative function of standard normal distribution. $\mu_{B_x }$ and $V_{B_x }$ such that $1-\Phi\left(\frac{q+\mu_{B_x }-  \widehat\mu_{E_x^{c}}}{\sqrt{    \widehat V_{E_x^{c}} -V_{B_x } }}\right) < r $ will explain away the estimated confounded treatment effect in meta-analysis. When $0<r<0.5$, a constant bias factor among studies, i.e., $V_{B_x }=0$, will lead to an upper bound of $  \widehat{p}(q)_x$, i.e., $$1-\Phi\left(\frac{q+\mu_{B_x }-  \widehat\mu_{E_x^{c}}}{\sqrt{    \widehat V_{E_x^{c}} -V_{B_x } }}\right)  \leq 1-\Phi\left(\frac{q+ \mu_{B_x }  -  \widehat \mu_{E_x^{c}}}{\sqrt{   \widehat  V_{E_x^{c}}}}\right), \   \ 0<r<0.5.$$ 
It means a constant bias factor, such that $ B_X \ge B_x^*$ where 
\begin{equation} \label{metab1}
 1-\Phi\left(\frac{q+ B_x^* - \widehat \mu_{E_x^{c}}}{\sqrt{    \widehat V_{E_x^{c}} }}\right) \equiv r \rightarrow
 B_x^* =\Phi^{-1}(1-r)  \widehat  V_{E_x^{c}}^{\frac{1}{2}}-q+ \widehat \mu_{E_x^{c}},
\end{equation}
and any other non-constant bias factors with mean larger or equal to $B_x^*$ can explain away the confounded treatment effect in meta-analysis. Thus, $B_x^*$ can be viewed as the \textit{minimal common constant bias factor} that reduce the probability of observing a CATE larger than a scientifically meaningful effect size $q$, to a threshold $r$.
  
For the apparently negative case, define $\Tilde{B }_x=  -B_x $; therefore $E_{k|x} =E_{k|x}^{c}+\Tilde{B}_{k|x} $ and $E_{k|x} \sim N(\mu_{E_{x}},V_{E_{x}})$ where $\mu_{E_{x}}=\mu_{E_x^c}+\mu_{\Tilde{B }_x}$ and $V_{E_{x}}=V_{E_x^c}-V_{\Tilde{B }_x}$. We define the probability of observing a CATE larger than $q$ as $p(q)_x=P\left(E_{k|x} <q\right)$. $p(q)_x$ can be expressed using $\mu_{\Tilde{B }_x }$, $ V_{\Tilde{B }_x}$, $ \mu_{E_x^{c}}$, and $ V_{E_x^{c}}$:\begin{equation*} 
{p}(q)_x= \Phi\left(\frac{q-\mu_{\Tilde{B }_x }- \mu_{E_x^{c}}}{\sqrt{   V_{E_x^{c}} -V_{\Tilde{B }_x} }}\right), \quad    V_{E_x^{c}} >V_{\Tilde{B }_x},  
\end{equation*} 
Similarly, when $0<r<0.5$, a constant bias factor among studies will lead to an upper bound of $\widehat{p}(q)_x$. Let $\Tilde{B}_x^*$ be the \textit{minimal common constant bias factor} that can reduce the probability of observing a conditional average treatment effects larger than a scientifically meaningful effect size $q$, to a threshold $r$ such that 
\begin{equation} \label{metab2}
\Phi\left(\frac{q-\Tilde{B}_x^*-\widehat\mu_{E_x^{c}}}{\sqrt{  \widehat V_{E_x^{c}} }}\right) \equiv r \rightarrow
 \Tilde{B}_x^*=q-\Phi^{-1}(r) \widehat V_{E_x^{c}}^{\frac{1}{2}}-\widehat\mu_{E_x^{c}}.
\end{equation}
 
Consider fixed $m_{k|1x}$, $m_{k|0x}$, $\theta_k$ and therefore a fixed bias factor across studies, meaning the unmeasured confounder has the same distribution and effect across all $K$ studies. In the apparently positive effect case, this means that if each of the $K$ studies suffers from unmeasured confounder with the same strength such that $ (m_{k|1x}-  m_{k|0x})\theta_k \ge \Phi^{-1}(1-r)  \widehat  V_{E_x^{c}}^{\frac{1}{2}}-q+ \widehat \mu_{E_x^{c}}$, the confounded treatment effect will be explained away. In the apparently negative effect case, it means that if $ (m_{k|0x}-  m_{k|1x})\theta_k \ge q-\Phi^{-1}(r) \widehat V_{E_x^{c}}^{\frac{1}{2}}-\widehat\mu_{E_x^{c}}$ in each of the $K$ studies, the confounded treatment effect will be explained away. 

Consider variant $m_{k|1x}$, $m_{k|0x}$, $\theta_k$ and therefore a variant bias factor across studies. In apparently positive effect case, any random bias factor with mean equal or larger than $\Phi^{-1}(1-r)  \widehat  V_{E_x^{c}}^{\frac{1}{2}}-q+ \widehat \mu_{E_x^{c}}$ can explain away the confounded treatment effect. In apparently negative effect case, any random bias factor with mean equal or larger than $q-\Phi^{-1}(r) \widehat V_{E_x^{c}}^{\frac{1}{2}}-\widehat\mu_{E_x^{c}}$ can explain away the confounded treatment effect.

Similar to the case of the single study, the robustness of the estimated confounded treatment effect can be measured by the minimal common constant bias factor. A small minimal common constant bias factor indicates that the estimated confounded treatment effect is not robust to unmeasured confounders.

Table \ref{Table:1} summarizes how to calculate the minimum (common constant) bias factor for a single study and for multiple studies (meta-analysis) under different scenarios. Once $B_x^*$ or $\Tilde{B}_x^*$ is calculated, one can determine the minimum effect of unmeasured confounders on response and difference between conditional mean of unmeasured confounders, that are sufficient to explain away the confounded treatment effect. 
\section{Interpretations}
\label{MM_interpretation} 
\subsection{Single study} 
We take a prospective clustered study, which focuses on identifying potential causal effect of breastfeeding on children's intelligence quotient (IQ), as an example to explain how to apply our method to assess the robustness of results from an observational study. \cite{Luby2016} examined relationship between breastfeeding and children's IQ. After adjusting for children's age, sex, history of depression, caregiver's highest level of education, and familial income-to-needs ratio, the authors found that the IQ of breastfed child was $5.49$ (95\% CI:  $0.75$ to $10.23$) higher than the IQ of non-breastfed child. The investigators did not control for maternal smoking during pregnancy while maternal smoking may be a confounder as it could associate with less breastfeeding as well as lower offspring IQ. We are worried this result is confounded by maternal smoking status, so we conduct a sensitivity analysis to assess how robust the estimate is to the unmeasured confounding: maternal smoking status. 

\begin{figure}[h]
\centering
\includegraphics[width=9cm]{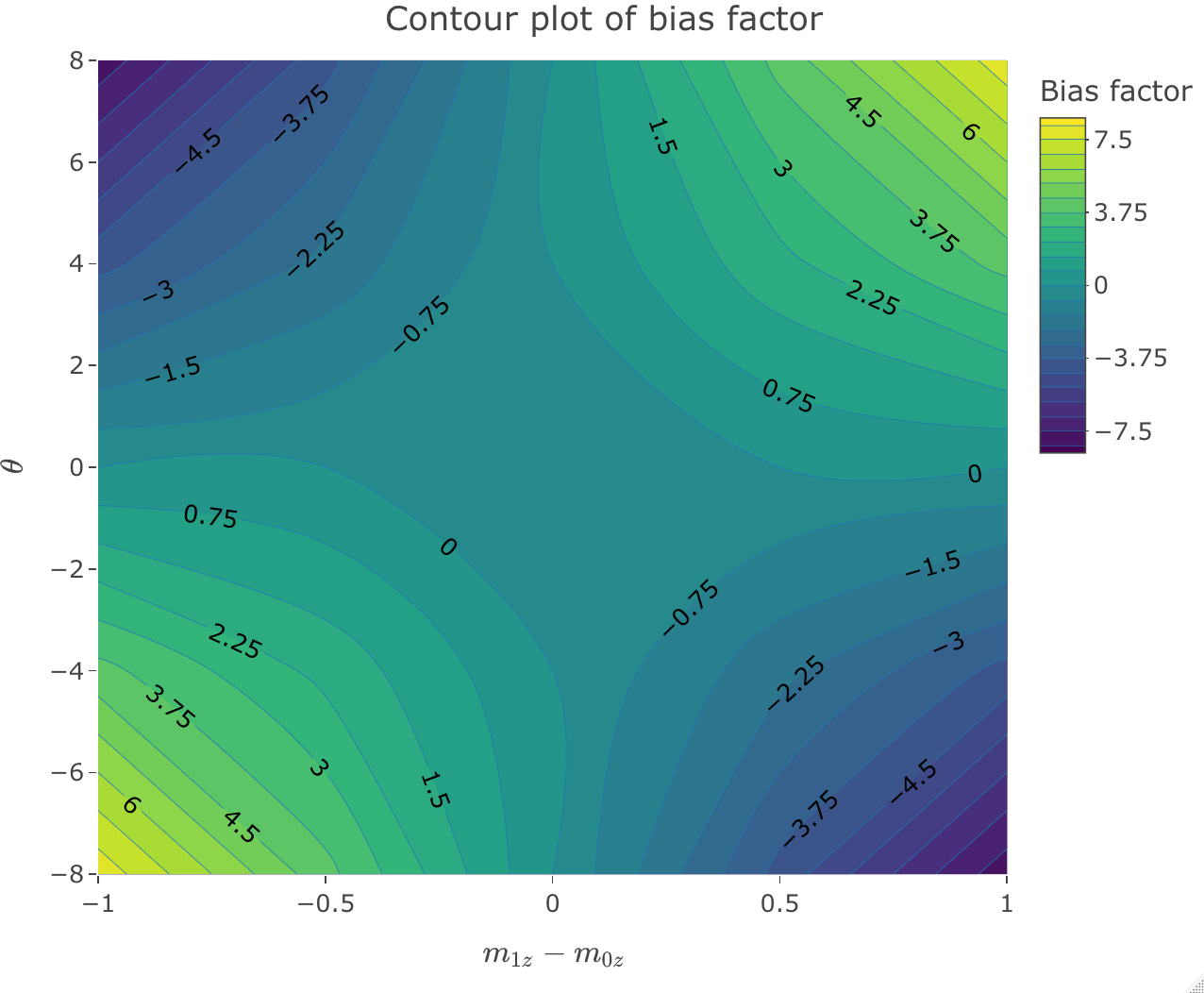}
\caption{The contour plot shows the relationship between $m_{1x}-  m_{0x}$ (x axis) and $\theta$ (y axis) and the bias factor. $ m_{0x}$: mean of unmeasured confounder in control group; $ m_{1x}$: mean of unmeasured confounder in treatment group; $\theta$: effect of unmeasured confounder on outcome
}
\label{Fig1}
\end{figure}

As the study indicated an apparently positive effect of breastfeeding on children's IQ, the minimal bias factor, $B_x^*$, that is able to explain away the confounded effect equals the lower bound of confidence interval, i.e., $B_x^*=0.75$, which means that any bias factor larger than $0.75$ can also explain away the confounded effect. The minimum bias factor $ 0.75$ can be decomposed into the product of two factors: the effect of maternal smoking on children's IQ adjusting for other covariates and the difference between $\mathbb{E}(U|A=1,X=x)$ and $\mathbb{E}(U|A=0,X=x)$. Our methods allow investigators to make statements of the following form: the confounded effect of breastfeeding on children’s IQ could be explained away if the difference between the prevalence of maternal smoking in breastfed children and non-breastfed children is larger than $0.25$ and the effect of smoking on IQ is greater than 3. There are also other decompositions that could explain away the confounded effect of breastfeeding on children’s IQ. A contour plot depicted in Figure \ref{Fig1} can be used to visualize the relationship between bias factors, difference between $\mathbb{E}(U|A=1,X=x)$ and $\mathbb{E}(U|A=0,X=x)$ (i.e., $m_{1x}-  m_{0x}$), and the effect of unmeasured confounder $\theta$. Any combination of $m_{1x}-  m_{0x}$ and $\theta$ in the contour plot that corresponds to a bias factor equal or larger than $0.75$ are able to explain away the confounded effect. Using both the information of the minimal bias factor and different combinations of sensitivity parameters that are able to explain away the confounded treatment effect, investigators can get a sense of how sensitive the conclusions are to potential unmeasured confounders.

\subsection{Multiple studies}
Applications of our methods for meta-analysis and single study are similar. Investigators can calculate the minimal common constant bias factor based on equations (\ref{metab1}) or (\ref{metab2}) by plugging in $\widehat V_{E_x^c}$, $\widehat\mu_{E_x^c}$, $r$, and $q$. If each of the studies suffers from an unmeasured confounder with the same distribution and effect on outcome that lead to a bias factor equal or greater than the minimal common constant bias factor, the confounded treatment effect from the meta-analysis can be explained away. Further, any non-fixed unmeasured confounder across studies with expectation equal or larger than minimal common constant bias factor can also explain away the confounded effect. Similar to single study, investigators can use our method to assess whether a set of sensitivity parameters  can explain away the confounded effect  and get a sense of how sensitive the conclusions are to potential unmeasured confounders.
 
We take a meta-analysis, which focuses on identifying potential causal effect of body mass index (BMI) in midlife on dementia, as an example to explain how to apply our method to assess the robustness of results from meta-analysis. \cite{Albanese2017} examined relationship between BMI in midlife and dementia by combining results, on relative risk (RR) scale, from 19 longitudinal studies on 589,649 participants (2,040 incident dementia cases) followed up for up to 42 years. The authors found that midlife (age 35 to 65 years) obesity (BMI $\ge$ 30) was associated with larger risk of dementia in late life (RR, 1.33; 95\% CI, 1.08 to 1.63). 
The study reported $\widehat\mu_{E_x^c}=\text{log} 1.33$ ($\widehat\mu_{E_x^c}$ is on log RR scale), while the study level summary measures reported from \cite{Albanese2017} can be used to obtain $\widehat V_{E_x^c}$. 
We estimated $\widehat V_{E_x^c}=0.08$ via methods illustrated in \cite{Borenstein:2007}. Note that the choice of $q$ and $r$ should be adapted based on the context and specific research questions, which means that different analysis can have difference choice on $q$ and $r$. In this example, let us consider a RR of 1.2 as a scientifically meaningful effect size, meaning $q=\text{log} 1.2$, and consider having at least 40\% chance to observe an effect size above $q$ for results to be of scientific interest, meaning $r=0.4$. 
 
As the meta-analysis indicated an apparently positive effect of BMI on dementia, we have the minimal common constant bias factor, which can reduce the probability of observing a conditional average treatment effects larger than a scientifically meaningful effect size $q$ by a threshold $r$, such that $$B_x^* =\Phi^{-1}(1-r)  \widehat  V_{E_x^{c}}^{\frac{1}{2}}-q+ \widehat \mu_{E_x^{c}}=\Phi^{-1}(1-0.4)  *\sqrt{0.08} -\text{log} 1.2+ \text{log} 1.33=0.17.$$ 
It means that any constant bias factor across studies such that $B_x  \ge 0.17$ and any other non-constant bias factors across studies with mean larger or equal to $0.17$ can explain away the confounded treatment effect in meta-analysis. The minimum common constant bias factor $0.17$ can be decomposed into the product of two factors: the effect of unmeasured confounder on dementia and the difference between the mean of unmeasured confounder in different BMI groups. Our methods allow investigators to make statement of the following form: the confounded effect of BMI on dementia could be explained away if an unmeasured confounder exists such that its mean difference between BMI groups is larger than 0.17 and its effect on dementia is greater than 1 in each of the studies. We can similarly refer to the contour plot in Figure \ref{Fig1} to visualize combinations of sensitivity parameters that corresponds to a bias factor equal or larger than 0.17 to explain away the confounded effect. Using both the information of the minimal common constant bias factor and different combination of sensitivity parameters that are able to explain away the confounded treatment effect, investigators can get a sense of how sensitive the conclusions are to potential unmeasured confounders.

\section{Simulation studies}
\label{MM_simulation}
We conducted simulation studies to assess the performance of our proposed methods; specifically, we examined the bias and the precision of the estimator $\widehat{E}_x$ in equation (\ref{bias1}) for a single study setting with continuous outcome, the estimator $\widehat{E}_x$ in equation (\ref{bias3}) for a single study setting with binary outcome and continuous unmeasured confounder, and estimator $\widehat{p}(q)_x$ in equation (\ref{pq}) for the meta-analysis settings.  
  
\subsection{Single Study - Continuous Outcome}
First, we assessed the performance of the estimator of $E_x$, $\widehat{E}_{x}=\widehat{\beta}_1^c+x\widehat{\beta}_3^c-B_x$. We assumed that $A$, $X$, and $U$ are dependent, which is a frequently encountered condition in observational studies.

Let the number of subjects be $J $ and the number of repeated measures be $I $ per subject. We simulated a normally distributed unmeasured confounder $U_{ij} \sim N(0,\sigma_u^2)$, one measured binary confounder $X_{ij} \sim Bin(0.5)$ if $U_{ij}<1$ and $X_{ij} \sim Bin(0.4)$ if $U_{ij} \ge 1$, and treatment assignment $A_{ij} \sim Bin(0.4)$ if $U_{ij}+X_{ij} < 2$ and $A_{ij} \sim Bin(0.5)$ if $U_{ij}+X_{ij} \ge 2$. We simulated random intercept $\zeta_j$ and the level-1 residuals $\epsilon_{ij}$ from the distributions $N(0,4)$ and $N(0,1)$, respectively. We proceeded to simulate the response variable $Y$ using the underlying true model described in (\ref{1}) setting $\beta_0=1$ and $\beta_2=3$ while varying the values of $\beta_1$, $\beta_3$, and $\theta$. $\widehat{E}_x$ can be calculated after we estimated the confounded coefficient $\beta_1^c$ and $\beta_3^c$ in (\ref{2}) using R package \texttt{lme4} (\citealp{Bates2015}). We simulated 1,000 such studies and inferred the estimated conditional average treatment effects described in equation (\ref{bias1}) under $X = 0$ and $X =1$. We repeated the simulations with different values of $I$, $J$, $\beta_1$, $\beta_3$, $\theta$, and $\sigma_u^2$. 
 
To measure the performance of our methods in estimating conditional causal treatment effect, we assessed the bias, standard error (SE) and coverage probability (CP) of 95\% confidence intervals of $\widehat{E}_x$. Results in Table \ref{Table:2} show that the estimated conditional average treatment effects, $\widehat{E}_x$, is unbiased and the standard error is relatively small. The coverage of 95\% confidence interval is nominal. The accuracy of the point estimate appears to be independent of $X$.

\subsection{Single Study - Binary Outcome}  Then, we assessed the performance of the estimator of $E_x$ in equation (\ref{bias3}) for a single study setting with binary outcome and continuous unmeasured confounder. Assuming 200 subjects and the 4 repeated measures per subject, $A$, $X$, and $U$ were simulated similarly as for continuous outcome. We simulated random intercept $\zeta_j$ and the level-1 residuals $\epsilon_{ij} $ from the distributions $N(0,\nu)$ and $N(0,1)$, respectively. We proceeded to simulate the response variable $Y$ using the underlying true model described in (\ref{binaryall}) setting $\beta_0=-4.5$,  $\beta_1= 1$,  $\beta_2=3$, and $\beta_3=-0.5$ while varying the value of $\theta$. 

As we have shown, equation (\ref{bias3}) holds when $\theta$, $\sigma_u$, true ICC, and confounded ICC are small, thus, in order to fully understand the relationship between bias of $\widehat {E}_{x}$, $\theta$, $\sigma_u$, and true ICC, we will estimate bias of $\widehat {E}_x$ under a set of different $\theta$, $\sigma_u$, and true ICC. Similarly, $\widehat{E}_x$ can be calculated after we estimated the confounded coefficient $\beta_1^c$ and $\beta_3^c$. We simulated 1,000 such studies. To measure the performance of our methods in estimating conditional causal treatment effect, we assessed the bias, standard error (SE) and coverage probability (CP) of 95\% confidence intervals of $\widehat{E}_x$. Results in Table \ref{Table:3} show that when $\theta$, $\sigma_u$, and true ICC is in range of simulated scenarios, our proposed estimators is unbiased and precise. There is no monotone relationship between accuracy of $ \widehat{E}_{x=0}$ and $\theta$, $\sigma_u$, or true ICC, as the approximation that allows equation (\ref{bias3}) to hold also depends on the confounded ICC which is related with all parameters.

\subsection{Multiple Studies} 
For multiple studies, we assessed the performance of the estimator of $p(q)_x$, $\widehat{p}(q)_x=1-\Phi\left(\frac{q+\mu_{B_x}-\widehat\mu_{E_x^c}}{\sqrt{  \widehat\nu_{E_x^c} -\nu_{B_x} }}\right)$. Let $K$ be the number of studies, $J$ be the subjects in each study and $I$ be the number of repeated measurements for each subject. Let $K \in \{15,30,100\}$, $\mathbb{E}(J) \in \{100,200\}$, and $I \in \{3 ,5 \}$. For each study, we simulate $J \sim Unif(\mathbb{E}(J)-50,\mathbb{E}(J)+50)$ to allow for between study variation in sample size. We simulated $\beta_{1k}$ and $\beta_{3k}$ from a bivariate normally distribution with $\mu_1=3$, $\mu_3=4$, $V_{11}=2$, $V_{33}=6$, and $V_{13}=0.05$, and simulated $\theta \sim N(5,0.01)$. Settings used in single study with continuous outcome were used to simulate $A$, $X$, $U$, $\zeta$, $\epsilon$, and $Y$ in each of the $K$ studies. Then we estimated the effect sizes in each of the $K$ studies and then fit the random effects meta-analysis using R package \texttt{meta} (\citealp{Schwarzer2015}). We repeated the process 1,000 times.

Results in Table \ref{Table:4} show that our proposed estimator $\widehat{p}(q)_x$ is unbiased and precise, as the SE are relatively small. The coverage of 95\% confidence interval is nominal. The proposed estimator performed well for all simulated conditions across various settings on the number of studies, subjects, and repeated measurements per subject.

\section{Discussion}
\label{MM_discussion}
In this study, we propose new sensitivity analysis methods to assess robustness of the treatment effect estimated from observational data against violation of the assumption of no unmeasured confounding. We developed methods for clustered outcomes (e.g., repeated measures per subject, patients nested within a clinic) of a single study and of multiple studies (meta-analysis) separately. Our methods enable researchers to estimate the minimal effect of the unmeasured confounder sufficient to explain away the confounded treatment effect, when applied after data analysis using the standard mixed effects modeling approach. If only an unmeasured confounder with strong effect can explain away the confounded effect, it means the observational study is robust to unmeasured confounding. The major advantage of our methods is   to avoid unrealistic assumptions, in addition to efficient computation and uncomplicated implementation.

Unlike previously developed methods, our methods can be easily extended to deal with multiple unmeasured confounders. Taking continuous outcome case as an example, equations (\ref{1}) to (\ref{bias1}) can be modified by including all unmeasured confounders, $U_1, \dots, U_L$, and the bias factor, $B_x$, can be derived similarly such that $B_x=\theta_1(m_{11x}-m_{10x})+ \dots + \theta_L(m_{L1x}-m_{L0x})$, where $\theta_l$ is the effect of $U_l$ on outcome and $m_{lax}=\mathbb{E}(U_l|A=a,X=x)$ for $l=1,\dots, L$, if we assume that unmeasured confounders are independent with each other. Our methods can be easily extended to deal with multiple treatments. To find the relationship between the CATE and confounded treatment effects in a study involving $S$ treatments, we need to obtain $m_{sx}=\mathbb{E}(U_{ij}| A_{ij}=s, X_{ij}=x)$ for $s=1, \dots, S$. The bias factor for each treatment effect will have the same form as that for two treatments. 

For multilevel data, for example, patients are nested within hospitals so unmeasured confounders could be at the patient or at the hospital level. In our derivations, we considered that the source of unmeasured confounders is at the patient level, so we incorporated $U_{ij}$ in the model. To assess the impact of unmeasured confounders at the hospital level, we may replace $U_{ij}$ with $U_j$ in the model and the same subsequent derivation follows. 

The minimal bias factor that can explain away the confounded result in a meta-analysis involves parameters $q$ and $r$ representing a scientifically meaningful effect size and the probability of observing a study that have an effect size larger than $q$, respectively. The choice of $q$ and $r$ should be adapted based on the context and specific research questions. For parameter $r$, our recommendation is to choose any value of $r$ less than 0.5 in order to keep equations (\ref{metab1}) and (\ref{metab2}) valid. Selection of $q$ should base on the research goals and the associated background knowledge. For example, if the goal of a cohort study is to test whether there is a treatment effect in a new physical therapy, the value of $q$ could be set to the treatment effect of the current protocol physical therapy.

While the focus of our proposed methods is on moving the confidence interval to include the null, we care not only about the statistical uncertainty of the estimate but also about the point estimate itself in practice. For this reason, researchers can also derive the bias factor that is able to move the point estimate to the null or even to reverse direction of the confounded effect. It is also possible to derive the minimal bias factor that can move a near null confounded treatment effect to a clinically meaningful level if it is the interest. No matter what target is set, a similar sensitivity analysis can be conducted once the (minimal) bias factor is calculated.

In practice, there could be violation against the assumption that the random effect and error term have constant variance in the confounded model, i.e., violation against the assumption that the variance of random effect and error term are independent of measured confounders and treatment. Given different relationship between unmeasured confounder, measured confounders, treatment and outcome, there could be different degree of violation. In our simulated scenarios, the data is generated in the way that this assumption is violated. The accurate and precise results of our simulation indicates that our methods are robust against the violation of the assumption that the random effect and error term have constant variance in the confounded model. 

Depending on the context, the same value of minimal bias factor could carry different interpretations. We suggest that for all observational studies whose goal is to investigate causal relationships, to report the minimal bias factor needed to explain away both the point estimate and confidence interval of the observed effect, and then interpret minimal bias factor and assess the robustness of the study to unmeasured confounding in its own context.

 \newpage 
\section*{Reference} 
\bibliographystyle{biorefs}
\bibliography{refs.bib}

\appendix


\section{Binary Outcome} \label{binaryoutcomeA}  
We establish a result that for binary outcome when $v$ is small, we have 
$$ 
 \text{logit}\{\mathbb{P}(Y_{ij}=1| A_{ij}, X_{ij},U_{ij} )\}\approx \beta_{0} + \beta_{1} A_{ij} +\beta_{2} X_{ij} + \beta_{3} A_{ij}  X_{ij}+\theta U_{ij}.
$$
$proof.$
\begin{equation*} 
\begin{split}
&\text{logit}\{\mathbb{P}(Y=1| A, X,U )\}\\
&=\text{log}\frac{ \int_{-\infty}^{\infty}\frac{\text{exp}(\beta_{0} + \beta_{1} A +\beta_{2} X + \beta_{3} A  X +\theta U +\zeta )}{1+\text{exp}(\beta_{0} + \beta_{1} A +\beta_{2} X + \beta_{3} A  X +\theta U +\zeta )}dF(\zeta | A, U,X) }  {1-\int_{-\infty}^{\infty}\frac{\text{exp}(\beta_{0} + \beta_{1} A +\beta_{2} X + \beta_{3} A  X +\theta U +\zeta )}{1+\text{exp}(\beta_{0} + \beta_{1} A +\beta_{2} X + \beta_{3} A  X +\theta U +\zeta )}dF(\zeta | A, U,X)} \\
&=\beta_{0} + \beta_{1} A +\beta_{2} X + \beta_{3} A  X+\theta U +\text{log}\frac{ \int_{-\infty}^{\infty}\frac{\text{exp}(\zeta )}{1+\text{exp}(\beta_{0} + \beta_{1} A +\beta_{2} X + \beta_{3} A  X +\theta U +\zeta )}dF(\zeta | A, U,X) }  {1-\int_{-\infty}^{\infty}\frac{\text{exp}(\beta_{0} + \beta_{1} A +\beta_{2} X + \beta_{3} A  X +\theta U +\zeta )}{1+\text{exp}(\beta_{0} + \beta_{1} A +\beta_{2} X + \beta_{3} A  X +\theta U +\zeta )}dF(\zeta | A, U,X)}.    
\end{split}
\end{equation*}  
Denote $h(A,X,U)=\int_{-\infty}^{\infty}\frac{\text{exp}(\zeta )}{1+\text{exp}(\Delta +\zeta )}dF( \zeta | A, X,U)$, where $\Delta=\beta_{0} + \beta_{1} A +\beta_{2} X + \beta_{3} A  X +\theta U $. Thus, $ \text{logit}\{\mathbb{P}(Y=1| A, X,U )\}=\beta_{0} + \beta_{1} A +\beta_{2} X + \beta_{3} A  X+\theta U +f(A,X,U),$ where $  f(A,X,U)=\text{log}\frac{h(A,X,U)}{1-\text{exp}(\Delta)*h(A,X,U)}$. 
If $v \rightarrow 0$, it means    
\begin{equation*}   
 h(A,X,U) \rightarrow  \frac{\text{exp}( \mathbb{E}(\zeta) )}{1+\text{exp}(\Delta  +\mathbb{E}(\zeta))} = \frac{1}{1+\text{exp}(\Delta  )}.
\end{equation*} 
Thus, when $v \rightarrow 0$
\begin{equation*}   
 f(A,X,U) \approx \text{log}\frac{\frac{1}{1+\text{exp}(\Delta  )}}{1-\frac{\text{exp}(\Delta)}{1+\text{exp}(\Delta  )}} = 0.
\end{equation*} 
Therefore, when $v \rightarrow 0$
\begin{equation*}   
 \text{logit}\{\mathbb{P}(Y=1| A, X,U )\}\approx \beta_{0} + \beta_{1} A +\beta_{2} X + \beta_{3} A  X+\theta U.
\end{equation*} 

\section{Binary Outcome and Binary Unmeasured Confounder} 
We establish a result that for binary outcome with presence of binary unmeasured confounder in the study, when both $v$ and $\theta$ are small, we have 
$$ \text{logit}\{\mathbb{P}(Y=1| A, X )\}=\beta_{0} + \beta_{1} A +\beta_{2} X + \beta_{3} A  X+\text{log}\{P_{Ax}  \text{exp}(\theta +\frac{\nu}{2 })+ (1-P_{Ax})  \text{exp}( \frac{\nu}{2 }) \}$$
$proof.$
\label{binaryoutcome}  
\begin{equation*} 
\begin{split}
&\text{logit}\{\mathbb{P}(Y=1| A, X )\} \\
&=\text{log}\frac{ \int_{-\infty}^{\infty}\int_{-\infty}^{\infty}\frac{\text{exp}(\beta_{0} + \beta_{1} A +\beta_{2} X + \beta_{3} A  X +\theta U +\zeta )}{1+\text{exp}(\beta_{0} + \beta_{1} A +\beta_{2} X + \beta_{3} A  X +\theta U +\zeta )}dF(U,\zeta | A, X) }  {1-\int_{-\infty}^{\infty}\int_{-\infty}^{\infty}\frac{\text{exp}(\beta_{0} + \beta_{1} A +\beta_{2} X + \beta_{3} A  X +\theta U +\zeta )}{1+\text{exp}(\beta_{0} + \beta_{1} A +\beta_{2} X + \beta_{3} A  X +\theta U +\zeta )}dF(U,\zeta | A, X)} \\
&=\beta_{0} + \beta_{1} A +\beta_{2} X + \beta_{3} A  X+\text{log}\frac{ \int_{-\infty}^{\infty}\int_{-\infty}^{\infty}\frac{\text{exp}(\theta U +\zeta )}{1+\text{exp}(\beta_{0} + \beta_{1} A +\beta_{2} X + \beta_{3} A  X +\theta U +\zeta )}dF(U,\zeta | A, X) }  {1-\int_{-\infty}^{\infty}\int_{-\infty}^{\infty}\frac{\text{exp}(\beta_{0} + \beta_{1} A +\beta_{2} X + \beta_{3} A  X +\theta U +\zeta )}{1+\text{exp}(\beta_{0} + \beta_{1} A +\beta_{2} X + \beta_{3} A  X +\theta U +\zeta )}dF(U,\zeta | A, X)}.    
\end{split}
\end{equation*}  
Denote $h(A,X)=\int_{-\infty}^{\infty}\int_{-\infty}^{\infty}\frac{\text{exp}(\theta U +\zeta )}{1+\text{exp}(\Delta +\theta U +\zeta )}dF(U,\zeta | A, X)$, where $\Delta=\beta_{0} + \beta_{1} A +\beta_{2} X + \beta_{3} A  X$. Thus, $ \text{logit}\{\mathbb{P}(Y=1| A, X )\}=\beta_{0} + \beta_{1} A +\beta_{2} X + \beta_{3} A  X+f(A,X), \ where\  f(A,X)=\text{log}\frac{h(A,X)}{1-\text{exp}(\Delta)*h(A,X)}$.
\begin{equation}  \label{binaryexpa}
\small \begin{split} 
h(A,X) &=\int_{-\infty}^{\infty}\int_{-\infty}^{\infty}\frac{\text{exp}(\theta U +\zeta )}{1+\text{exp}(\Delta +\theta U +\zeta )}dF(U,\zeta | A, X)\\
&=\int_{-\infty}^{\infty} \text{exp}(\theta U)\int_{-\infty}^{\infty}\frac{ \text{exp}(\zeta )}{1+\text{exp}(\Delta +\theta U +\zeta )}\frac{1}{\sqrt{2\pi\nu}} \text{exp}(\frac{-\zeta^2}{2\nu})d\zeta dF( U | A, X)\\
&=\int_{-\infty}^{\infty} \text{exp}(\theta U)\int_{-\infty}^{\infty}\frac{ \text{exp}(t \sqrt{\nu} )}{1+\text{exp}(\Delta +\theta U +t\sqrt{\nu} )}\frac{1}{\sqrt{2\pi\nu}} \text{exp}(\frac{-t^2}{2 })\sqrt{\nu} dtdF( U | A, X),\ t=\frac{\zeta}{\sqrt{\nu}} \\
&=\int_{-\infty}^{\infty} \text{exp}(\theta U)\frac{1}{\sqrt{2\pi }}\int_{-\infty}^{\infty}\frac{ 1}{1+\text{exp}(\Delta +\theta U +(s+\sqrt{\nu})\sqrt{\nu} )} \text{exp}(\frac{-s^2}{2 })\text{exp}(\frac{\nu}{2 }) ds dF( U | A, X),\ s=t-\sqrt{\nu} \\
&=\int_{-\infty}^{\infty} \frac{\text{exp}(\theta U+\frac{\nu}{2 })}{\sqrt{2\pi }}\int_{-\infty}^{\infty}\frac{ 1}{1+\text{exp}(\Delta +\theta U + s \sqrt{\nu}+\nu )} \text{exp}(\frac{-s^2}{2 }) ds dF( U | A, X) \\
&= P_{Az} \frac{\text{exp}(\theta +\frac{\nu}{2 })}{\sqrt{2\pi }}\int \frac{ 1}{1+\text{exp}(\Delta +\theta  + s \sqrt{\nu}+\nu )} \text{exp}(\frac{-s^2}{2 }) ds   + \\
&(1-P_{Ax}) \frac{\text{exp}( \frac{\nu}{2 })}{\sqrt{2\pi }}\int \frac{ 1}{1+\text{exp}(\Delta  + s \sqrt{\nu}+\nu )} \text{exp}(\frac{-s^2}{2 }) ds,
\end{split}
\end{equation} 
where $P_{AX}=P(U_{ij}=1 | A_{ij}, X_{ij})$. 
\begin{equation}  \label{binaryexpb}
\begin{split}
&1-\text{exp}(\Delta)*h(A,X=x) \\
&=P_{Ax}+ (1-P_{Ax})-\text{exp}(\Delta)*h(A,X=x)\\
&= P_{Ax}\int_{-\infty}^{\infty}1-\frac{ \text{exp}(\Delta +\theta  +\zeta)}{1+\text{exp}(\Delta +\theta  +\zeta)}  dF( \zeta | A, X=x) +\\
& (1-P_{Ax})\int_{-\infty}^{\infty}1-\frac{ \text{exp}(\Delta   +\zeta)}{1+\text{exp}(\Delta  +\zeta)}  dF( \zeta | A, X=x)
\\
&= P_{Ax}\int_{-\infty}^{\infty} \frac{ 1}{1+\text{exp}(\Delta +\theta  +\zeta)} \frac{1}{\sqrt{2\pi\nu}} \text{exp}(\frac{-\zeta^2}{2\nu}) d \zeta +\\
& (1-P_{Ax})\int_{-\infty}^{\infty} \frac{ 1}{1+\text{exp}(\Delta  +\zeta)} \frac{1}{\sqrt{2\pi\nu}} \text{exp}(\frac{-\zeta^2}{2\nu}) d \zeta \\
&= \frac{P_{Ax}}{\sqrt{2\pi }}\int_{-\infty}^{\infty} \frac{ 1}{1+\text{exp}(\Delta +\theta  +t\sqrt{\nu})}   \text{exp}(\frac{-t^2}{2 }) d t +\\
& \frac{(1-P_{Ax})}{\sqrt{2\pi }}\int_{-\infty}^{\infty} \frac{ 1}{1+\text{exp}(\Delta  +t\sqrt{\nu})}  \text{exp}(\frac{-t^2}{2 }) d t ,\ t=\frac{\zeta}{\sqrt{\nu}}. 
\end{split}
\end{equation} 

Based on the results from equation (\ref{binaryexpa}) and equation (\ref{binaryexpb}):
\begin{equation*}  \label{binaryexpF}
\small \begin{split}
&f(A,X=x)=\text{log}\frac{P_{Ax}  \text{exp}(\theta +\frac{\nu}{2 }) \int \frac{ 1}{1+\text{exp}(\Delta +\theta  + s \sqrt{\nu}+\nu )} \text{exp}(\frac{-s^2}{2 }) ds   + (1-P_{Ax})  \text{exp}( \frac{\nu}{2 }) \int \frac{ 1}{1+\text{exp}(\Delta  + s \sqrt{\nu}+\nu )} \text{exp}(\frac{-s^2}{2 }) ds}  { P_{Ax} \int_{-\infty}^{\infty} \frac{ 1}{1+\text{exp}(\Delta +\theta  +t\sqrt{\nu})}   \text{exp}(\frac{-t^2}{2 }) d t +  (1-P_{Ax}) \int_{-\infty}^{\infty} \frac{ 1}{1+\text{exp}(\Delta  +t\sqrt{\nu})}  \text{exp}(\frac{-t^2}{2 }) d t}.
\end{split}
\end{equation*} 
When both $\theta$ and $\nu$ are small, we have
\begin{equation*}  
f(A,X=x) \approx \text{log}\{P_{Ax}  \text{exp}(\theta +\frac{\nu}{2 })+ (1-P_{Ax})  \text{exp}( \frac{\nu}{2 }) \}.
\end{equation*}

\section{Binary Outcome and Continuous Unmeasured Confounder} 
\label{binaryoutcome2}  
We establish a result that for binary outcome with presence of continuous unmeasured confounder in the study, when both $v$, $\theta$, and $\sigma_u$ are small, we have 
$$ \text{logit}\{\mathbb{P}(Y=1| A, X )\}=\beta_{0} + \beta_{1} A +\beta_{2} X + \beta_{3} A  X+  \theta\mu_{AX} +\frac{\theta^2\sigma_u+\nu}{2 }$$
$proof.$
Denote $l=\theta U+\zeta$, as we assume that $U \perp \zeta \ |\  A,X$, thus $l= \theta U+ \zeta \sim N(\mu_{l},\sigma_l) $, where $\mu_{l}=\theta\mu_{l}$ and $\sigma_l=\theta^2\sigma_u+\nu$. We have,
\begin{equation*} \label{normal1} 
h( A, X )  = \frac{\text{exp}(\mu_{l} +\frac{\sigma_l}{2 })}{\sqrt{2\pi }}\int \frac{ 1}{1+\text{exp}(\Delta +\mu_{l} + m \sqrt{\sigma_l}+\sigma_l )} \text{exp}(\frac{-m^2}{2 }) dm,\ m=\frac{l-\mu_{l}}{\sqrt{\sigma_l}}- \sqrt{\sigma_l}; 
\end{equation*} 
\begin{equation*} \label{normal2} 
1-\text{exp}(\Delta)*h( A, X )  = \frac{1}{\sqrt{2\pi }}\int \frac{ 1}{1+\text{exp}(\Delta +\mu_{l} + n \sqrt{\sigma_l}  )} \text{exp}(\frac{-n^2}{2 }) dn,\ n=\frac{l-\mu_{l}}{\sqrt{\sigma_l}}; 
\end{equation*}  
\begin{equation*}  \label{normalF}
\begin{split}
&f(A,X=x)=\text{log}\frac{\frac{\text{exp}(\mu_{l} +\frac{\sigma_l}{2 })}{\sqrt{2\pi }}\int \frac{ 1}{1+\text{exp}(\Delta +\mu_{l} + m \sqrt{\sigma_l}+\sigma_l )} \text{exp}(\frac{-m^2}{2 }) dm}  { \frac{1}{\sqrt{2\pi }}\int \frac{ 1}{1+\text{exp}(\Delta +\mu_{l} + n \sqrt{\sigma_l}  )} \text{exp}(\frac{-n^2}{2 }) dn}.
\end{split}
\end{equation*} 
When $\sigma_l$ is small,
\begin{equation*} \label{normal2}
f(A,X=x) \approx \mu_{l} +\frac{\sigma_l}{2 }.
\end{equation*}  

\newpage 

\begin{table}[h]
\caption{Calculating minimum (common constant) bias factor for single study and multiple studies (meta analysis). 
}\label{Table:1}
\centering
\begin{threeparttable} 
\begin{tabular}{lccrr}
\hline
& \text{Single study}&  \text{Multiple studies (meta-analysis)} \\  
\hline
\text{Apparently positive treatment effect}&$B_x^* =lb $ &$ B_x^* =\Phi^{-1}(1-r) \widehat V_{E_x^{c}}^{\frac{1}{2}}-q+\widehat\mu_{E_x^{c}}$ \\ 
\text{Apparently negative treatment effect}&$\Tilde{B}_x^*\ =-ub\tnote{2}$ &$\Tilde{B}_x^*=q-\Phi^{-1}(r) \widehat V_{E_x^{c}}^{\frac{1}{2}}-\widehat\mu_{E_x^{c}}$\\\hline 
\end{tabular}
$B_x^*$: minimum (common constant) bias factor in study with apparently positive treatment effect; $\Tilde{B}_x^*$: minimum (common constant) bias factor in study with apparently negative treatment effect; $lb$: lower bound of confidence interval; $ub$: upper bound of confidence interval; $\widehat\mu_{E_x^{c}}$: estimated grand mean of confounded treatment effect in meta analysis; $\widehat V_{E_x^{c}}^{\frac{1}{2}}$: estimated between study variance of confounded treatment effect in meta analysis; $q$: a scientifically meaningful effect size; $r$: threshold of probability of observing a conditional average treatment effects larger than a scientifically meaningful effect size $q$ for results to be of scientific interest
\end{threeparttable}
\end{table}

\begin{table}[h]
\centering
\caption{Bias, standard error (SE) and coverage probability (CP) of 95\% confidence intervals of the estimated conditional average treatment effects, $\widehat{E}_x$ conditional on $X=0$ and $X=1$ in single study with continuous outcome.
\label{Table:2}}
\begin{threeparttable}
\begin{tabular}{cccccccccccccc}
\hline
 &  &  &  & & &\multicolumn{3}{c}{$X=0$} &  & \multicolumn{3}{c}{$X=1$} \\ \cline{7-9} \cline{11-13}  
 $J$ & $I$  &$\beta_1$  & $\beta_3$ & $\theta$ & $\sigma_u^2$&Bias &SE & CP & &Bias & SE &CP \\ \hline 
 $50$ & $3$  &$-1$  & $1$ & $0.5$ & $0.25$&0.018 &0.281 & 0.948 & &0.009 & 0.285 &0.956 \\ 
 $100$ & $3$  &$-1$  & $1$ & $0.5 $ & $0.25$& 0.03 &0.157 & 0.952 & &0.017 & 0.214 &0.951 \\ 
 $100$ & $8$  &$-1$  & $1$ & $0.5 $ & $0.25$& 0.006 &0.109 & 0.951 & &0.002 & 0.115 &0.948 \\  
 $100$ & $3$  &$-1$  & $1$ & $-0.5 $& $0.25$& 0.004 &0.203 & 0.94 & & 0.002 & 0.199 &0.96\\   
 $100$ & $3$  &$1$  & $1$ & $0.5 $ & $0.25$&0.01 &0.195 & 0.952 & &0.009 & 0.212 &0.954 \\  
 $100$ & $3$  &$1$  & $-1$ & $0.5 $ & $0.25$& 0.001 &0.159 & 0.954 & & 0.005 & 0.210 &0.959 \\ 
 $100$ & $3$  &$-1$  & $1$ & $0.5 $ & $1$& 0.002 &0.155 & 0.969 & &  0.020 & 0.199 &0.967   \\ \hline
\end{tabular} 
$J$: size of level-1 unit (e.g., clusters); $I$: size of level-2 unit (e.g., subjects nested in clusters); $\boldsymbol \beta $: regression parameters; $\theta$: effect of unmeasured confounder on outcome; $\sigma_u^2$: variance of unmeasured confounder
\end{threeparttable}
\end{table}

\begin{table}[h]
\centering
\caption{Bias, standard error (SE) and coverage probability (CP) of 95\% confidence intervals of the estimated conditional average treatment effects, $\widehat{E}_x$ conditional on $X=0$ and $X=1$ in single study with binary outcome.
\label{Table:3}}
\begin{threeparttable}
\begin{tabular}{cccccccccc}
\hline
& & & \multicolumn{3}{c}{$X=0$} &  & \multicolumn{3}{c}{$X=1$} \\ \cline{4-6} \cline{8-10}  
$\theta$& ICC & $\sigma_u^2$ & Bias &SE & CP & &Bias & SE &CP \\ \hline 
-0.5&0.15&0.25&  0.001  &0.987 & 0.983  &&  0.025 &0.289 &0.960    \\
-0.5&0.15&1& 0.047 &0.707 &0.981   &&  0.031 & 0.278& 0.957   \\
-0.5&0.15&2.25& 0.034 &0.670 & 0.968  &&  0.048 &0.376 &0.959    \\
-0.5&0.25&0.25& 0.028 &0.637 &0.974   &&0.019   &0.301 &0.952    \\
-0.5&0.25&1& 0.039 &0.659 &0.972   && 0.034  &0.292 &0.953    \\
-0.5&0.25&2.25& 0.033 &0.588 &0.967   && 0.033  &0.287 &0.967    \\ 
-0.5&0.35&0.25& 0.045 & 0.586&0.963   &&0.008   &0.321 &0.956    \\
-0.5&0.35&1& 0.032 &0.575 & 0.968  &&0.005   &0.304 &0.954    \\
-0.5&0.35&2.25&0.079  &0.579 &0.954   &&0.056   &0.307 &0.951    \\ 
0.5&0.25&0.25& 0.012 &0.649 & 0.969  &&0.004   &0.302 &0.953    \\
0.5&0.25&1& 0.047&0.645 &0.971  &&0.034   &0.289 &0.954    \\
0.5&0.25&2.25&0.057  &0.602 &0.968   &&0.038   &0.275 &0.958    \\  \hline
\end{tabular} 
$\theta$: effect of unmeasured confounder on outcome; ICC: intraclass correlation coefficient; $\sigma_u^2$: variance of unmeasured confounder
\end{threeparttable}
\end{table}

\begin{table}[h]
\centering
\caption{Bias, Standard Error (SE) and coverage probability (CP) of 95\% confidence intervals  of the estimated probability of observing a conditional average treatment effects larger than $q$, $\widehat{p}(q)_x$, conditional on $X=0$ and $X=1$ in meta analysis.
\label{Table:4}}
\begin{threeparttable}
\begin{tabular}{cccccccccc}
\hline
 &  &  & \multicolumn{3}{c}{X=0} &  & \multicolumn{3}{c}{X=1} \\ \cline{4-6} \cline{8-10} 
 $K$&$\mathbb{E}(J)$&$I$&Bias &SE &CP &  &Bias & SE&CP  \\ \hline
 15&100&3&0.048& 0.125& 0.956& &0.001&0.126 &0.962   \\
 15&100&5&0.094&0.124& 0.937& &0.006&0.126 &0.973   \\
 15&200&3&0.079&0.124&0.940 & &0.025&0.125 &0.96   \\
 15&200&5&0.05&0.125& 0.964 & &0.017&0.124 &0.966    \\\hline
 30&100&3&0.044&0.089& 0.962& &0.008&0.090  &0.972  \\
 30&100&5&0.056&0.089&0.941 & &0.031&0.090 &0.972  \\
 30&200&3&0.046&0.088&0.936  & &0.015&0.090  &0.98  \\
 30&200&5&0.009&0.090&0.979  & &0.030&0.091  &0.969 \\\hline
 100&100&3&0.012&0.049&0.985 & &0.009&0.048   &0.985 \\
 100&100&5&0.002&0.048&0.987 & &0.017&0.049  &0.981  \\
 100&200&3& 0.004&0.049&0.972 & &0.009&0.049 &0.989  \\
 100&200&5& 0.003&0.048& 0.989& &0.005&0.048 &0.989\\ \hline
\end{tabular}
$K$: size of studies included in meta analysis; $\mathbb{E}(J)$: average size of level-1 unit (e.g., clusters); $I$: size of level-2 unit (e.g., subjects nested in clusters);
\end{threeparttable}
\end{table}

\end{document}